# Crop-specific Optimization of Bifacial PV Arrays for Agrivoltaic Food-Energy Production: The Light-Productivity-Factor Approach

Muhammad Hussnain Riaz[1], Hassan Imran[1], Habeel Alam[1], Muhammad Ashraful Alam[2], and Nauman Zafar Butt[1]

[1]Department of Electrical Engineering, Lahore University of Management Science, Lahore, Pakistan
[2]Electrical and Computer Engineering Department, Purdue University, West Lafayette, IN, USA

*Abstract*— Agrivoltaics (AV) is an emerging technology having symbiotic benefits for food-energy-water needs of the growing world population and an inherent resilience against climate vulnerabilities. An agrivoltaic system must optimize sunlight-sharing between the solar panels and crops to maximize the food-energy yields, subject to appropriate constraints. Given the emerging diversity of monofacial and bifacial farms, the lack of a standardized crop-specific metric (to evaluate the efficacy of the irradiance sharing) has made it difficult to optimize/assess the performance of agrivoltaic systems. Here we introduce a new metric – light productivity factor (LPF) – that evaluates the effectiveness of irradiance sharing for a given crop type and PV array design. The metric allows us to identify optimal design parameters including the spatial PV array density, panel orientation, and single axis tracking schemes specific to the PAR needs of the crop. By definition, LPF equals 1 for PV-only or crop-only systems. The AV systems enhances LPF between 1 and 2 depending on the shade sensitivity of the crop, PV array configuration, and the season. While traditional fixed-tilt systems increase LPF significantly above 1, we find LPF is maximized at 2 for shade-tolerant crops with a solar farm based on single axis sun tracking scheme. Among the fixed tilt systems, East/West faced bifacial vertical solar farms is particularly promising because it produces smallest variability in the seasonal yield for shade sensitive crops, while providing LPF comparable to the standard N/S faced solar farms. Additional benefits include reduced soiling and ease of movement of large-scale combine-harvester and other farming equipment.

## I. Introduction

ALONG with many positive aspects of rapidly growing photovoltaic (PV) installations, the rapid growth of ground mounted PV farms raises important environmental concerns including the land use conflict with agriculture, adverse impact on ecosystem processes, and loss of biodiversity [1-3]. Minimizing these environmental challenges is crucial for the desired techno-ecological growth of solar PV for increasing world population especially in regions that are highly susceptible to heat stress, drought, and climate change [1]. Recently, an innovative approach of agrivoltaic (AV) farming in which PV and agriculture are collocated has been demonstrated that offers a range of symbiotic benefits including dual food-energy production, reduced water budget, and resistance to climate effects such as excess heat and drought [4-8].

Unlike a PV-only or crop-only farms, AV farms are seldom optimized for a specific metric (e.g., LCOE); instead, one adopts an empirical design-of-experiment approach to study the overall farm yield at a given location. A review of the existing AV systems shows that the PV arrays are typically installed 4 to 7 meters above the crop level and at a lower density to reduce shading and improve sunlight-sharing between PV modules and crops. There have recently been several field experiments and some modeling studies to assess AV approach for a range of crops including lettuce, wheat, corn, tomatoes, cucumber, and peppers under standard and reduced spatial density of PV arrays [4, 8-13]. These initial studies predict that although crop yield can vary under the partial shading of an AV farm as compared to that in open sun, AV yields are not reduced significantly and, in some cases, can even exceed to that in open sun when the array density is carefully selected. Moreover, it has been proposed that cumulative radiation available for crops could be best manipulated through dynamic tilt control of PV panels through custom tracking schemes [13]. Marrou et al. [11] showed that the change in the intensity of radiation with respect to open sun condition is the dominant factor for the relative crop yield in an AV farm, although other microclimate parameters, such as the temperature and humidity, may also vary under the AV shades [14, 15]. In essence, although the importance of managing the sunlight balance between solar modules and crops is well-established, there is no systematic approach to a-priori optimize its design.

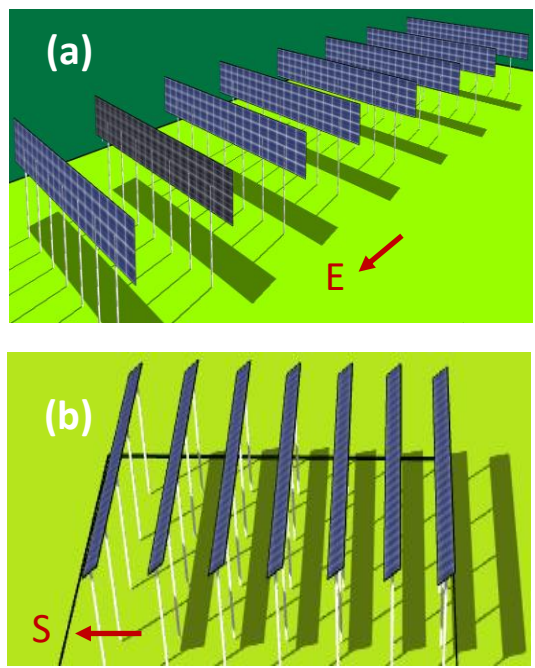

Fig.1. (a) E/W faced vertical panels, (b) N/S faced panels at an optimal fixed tilt.

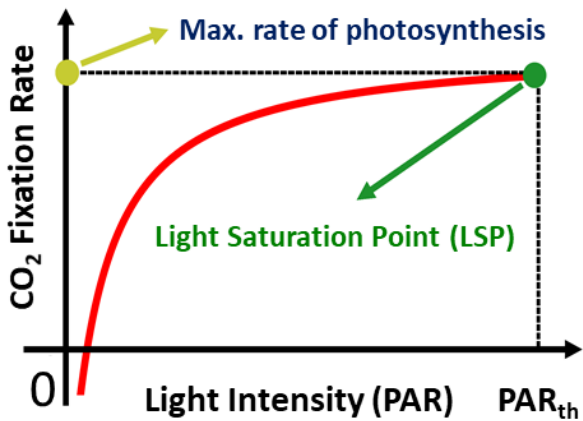

Fig.2. Typical response of the rate of photosynthesis as a function of sunlight intensity.

Any systematic approach for design optimization must rely on a metric to characterize the farm productivity for a given balance of sunlight between PV modules and crops. Unfortunately, the community is yet to define a suitable metric that can be used *in the design phase*. For example, in a simple approach, the sunlight sharing in AV could be characterized by the amount intercepted by the panels and the photosynthetically active radiation (PAR) available to the crops under the PV arrays. Recently, Hussnain et al [10] has explored the relative performance of N/S faced fixed tilt PV arrays vs. E/W faced vertical bifacial PV arrays (see Fig. 1) based on this simple sunlight sharing model. While this method is simple, incident PAR under panels is a crop-independent parameter and cannot be used for a crop-specific optimization. On the other hand, sophisticated mechanistic crop models have been used to predict the crop yield as a function of shading as well as other parameters [6, 9]. The physiological response of crops in AV systems is however a topic of active research and the validity of various mechanistic crop models under artificial shading of solar modules await field validation [16]. A simpler yet crop-specific approach is needed to optimize the agrivoltaic design of PV arrays and the choice of suitable crops.

In this paper, we introduce a crop-specific metric that can indicate potential AV performance for a given PV array design. The approach is based on evaluating how an AV system alters the availability of useful PAR ($PAR_u$) at the crop level, as compared to that in an open field. The $PAR_u$ is the daily integrated PAR that contributes to the crop's photosynthesis process. It is well-known that the rate of photosynthesis in plants is proportional to the PAR intensity up to a certain threshold value that is known as the light saturation point or threshold PAR ($PAR_{th}$), see Fig. 2. Increasing PAR intensity above $PAR_{th}$ does not increase the photosynthesis rate. For AV systems, it is therefore useful to evaluate how the shading under the panels varies the $PAR_u$ across various seasons for a given crop. When $PAR_u$ is combined with the intercepted solar irradiation by the solar modules, net productivity of sunlight for food-energy can be obtained. We quantify this in terms of a new metric called the Light Productivity Factor (*LPF*). As a correlated metric to experimentally reported metric known as the Land Equivalent Ratio (*LER*) (which indicates food-energy performance based on the *measured* crop yield and the PV output), *LPF* would enable a crop-specific optimization of the PV array density, orientation, and the mobile tilt algorithms at the design phase. Using this approach and as an illustrative example, we evaluate various AV design configurations for Lahore (31.5204° N, 74.3587° E) using typical meteorological data. In particular, we explore: (i) Farm productivity for fixed tilt bifacial PV arrays that include the standard N/S faced optimally tilt panels vs. E/W faced vertical bifacial panels; (ii) Farm productivity for the standard vs. non-standard single axis tracking schemes using mobile tilt bifacial panels; (iii) Crop dependence and seasonal variations in the farm productivity for fixed and mobile tilt PV systems; and (iv) Performance evaluation of various PV array configurations for a given constraint on the crop yield.

Three representative crops, *i.e.,* lettuce, turnip, and corn are considered to cover a broad range of $PAR_{th}$ while a variety of monofacial and bifacial panel configurations are explored to illustrate the relative design trade-offs. Although other practical considerations such as the required elevation of the arrays for farm machinery movement, climatic stresses on crops, and soiling power loss can influence the choice of a suitable PV array design for agrivoltaics, here we will focus exclusively on the food-energy productivity of the farm for unstressed crops. This paper is divided into four sections. In Section II, we describe in detail the modeling methodology. Simulation and modeling results are discussed in Section III, and the key conclusions of the paper is summarized in Section IV.

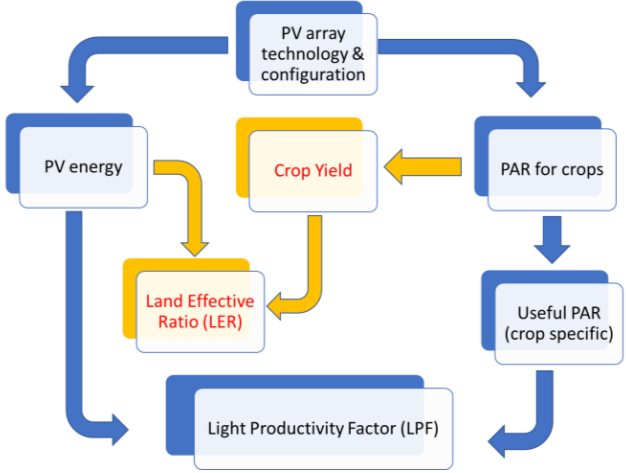

Fig. 3. Comparison of the modeling frameworks based on PAR, LER, and LPF approaches.

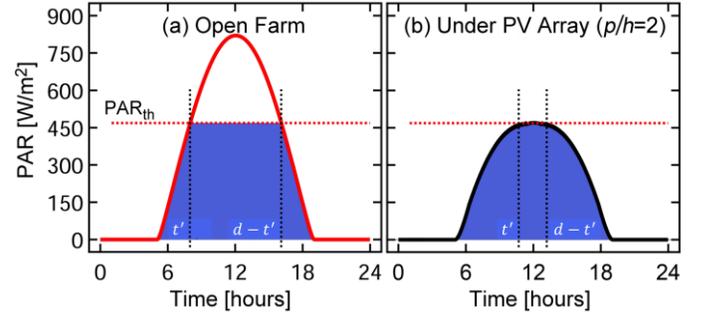

Fig. 4 PAR incident on the crop across a day for open (left) and under full density PV array (right). $t'$ and $d - t'$ are the times before and after noon, respectively, when the incident $PAR$ reaches to $PAR_{th}$ for the crop. $d$ is the duration of sunlight hours.

## II. MODELING APPROACH

### A. An integrated model for PV energy yield and PAR

A two-dimensional (2-D) model, appropriate for typical large scale commercial farms arranged in very long rows, is used to compute the solar irradiance interception along the height of the panels, the generated PV energy, and the incident PAR under the panels as a function of space and time per unit farm area. The length of the PV rows and the total number of rows are assumed large so that edge effects and variations along the dimension of PV rows are negligible. The 2-D model can accurately predict the system performance (except close to the edges) with a much-reduced analytical complexity as compared to 3-D approach [17].

The details of the analytical 2-D model have been reported elsewhere [10]. Briefly, the solar irradiance is evaluated as a function of time during the day for the location specified by its latitude and longitude using Sandia's PVLib library [18] to calculate the Sun's zenith and azimuth angles. The Global Horizontal Irradiance (GHI) is first calculated using Haurwitz clear sky model implemented in PVLib and is then renormalized based on the typical meteorological data from NASA Surface meteorology and Solar Energy database [19]. The GHI is then split into DNI and DHI components using Orgill and Hollands model [20].

To compute the PV energy and incident PAR underneath the PV arrays, the spatial-temporal interception of the direct and diffuse sunlight along the height of panels and along horizontal plane underneath the PV arrays is evaluated using the view factors approach [10, 21-23]. The albedo collection at the panels is similarly modeled. An accurately implemented view-factor approach provides the same accuracy as the ray-tracking approach, but at a much-reduced computational cost. The approach has been validated with published field data for both standard and agrivoltaic farms [10, 21]. The overall modeling framework is illustrated in Fig. 3.

### B. PV Array configurations

The farm productivity is analyzed for PV arrays with varying spatial density measured relative to optimized PV-only farm. Full (standard) array density is defined when row to row pitch ($p$) is twice as that of the panel height ($h$), i. e., $p/h = 2$. Similarly, half-density farm is defined by ($p/h=4$). For the performance comparison, the tracking schemes are compared with fixed-tilt N/S facing monofacial and E/W facing vertical bifacial farms. The definition of full, half, or quarter densities farms are taken from standard AV literature.

For sun-tracking tilt, single axis horizontal tracking for the E/W facing bifacial PV modules is considered with various tracking schemes. The standard tracking (ST) scheme is defined when modules are oriented normal to direct sunbeam. On the other hand, for reverse/anti-tracking (RT) scheme, modules are oriented parallel to the direct beam. For RT, although direct light shadowing is absent, the diffused sunlight under the panels is blocked/masked and the overall PAR intensity is lower compared to the open field. The customized tracking (CT) scheme switches between ST around noon and RT closer to sunrise and sunset. It implements ST for $n$ number of hours symmetrically before and after the midday. For other times during the day, RT is implemented.

### C. Useful PAR Yield and Light Productivity Factor

It is well established that each crop has a threshold PAR, $PAR_{th}$, above which the photosynthesis rate saturates [19]. The PAR available to the crops reduces under PV arrays as a function of the array density and the mobile tilt algorithm.

Under open sun and clear sky conditions, the time varying PAR ($PAR_0$) can be expressed as:

$$PAR_o(t) = C_{AM1.5} \times GHI\ (t) \qquad (1)$$

where $C_{AM1.5} \approx 0.51$ is the ratio of the integrated PAR (400nm – 700nm) to the integrated AM1.5 sunlight spectrum. For an agrivoltaic system, the spatially average, time varying PAR on a horizontal plane at any height under the panels can be evaluated as:

$$PAR_{AV}(t) = C_{AM1.5} \times (GHI(t) - G_{pV}(t)) \quad (2)$$

where $G_{pV}$ is the sum of spatially averaged direct and diffused irradiation on a horizontal plane that is intercepted or masked by solar modules, respectively.

By integrating the time-varying PAR over the entire day while imposing an upper limit of $PAR_{th}$ for a given crop, we can obtain a crop-specific useful daily PAR:

$$PAR_{u,open} = 2\int_0^{t'} PAR_0(t)\,dt + (d - 2t') \times PAR_{th} \quad (3a)$$

$$PAR_{u,AV} = 2\int_0^{t'} PAR_{AV}(t)\,dt + (d - 2t')PAR_{th} \quad (3b)$$

where $PAR_{u,open}$ and $PAR_{u,AV}$ are daily useful $PAR$ for open sun and under solar panels respectively, $t'$ and $d - t'$ are the time instances when the incident $PAR$ reaches to $PAR_{th}$ in the morning and afternoon, respectively, and $d$ is the duration of sunlight hours for the day (See Fig. 4). $PAR_{th}$ for lettuce, turnip, and corn are taken to be 213 $W/m^2$ (25 $kLx$), 469 $W/m^2$ (55 $kLx$) and 685 $W/m^2$ (80 $kLx$), respectively [20].

Let us now define a yield ratio for the useful daily PAR:

$$Y_{PAR} = \frac{PAR_{u,AV}}{PAR_{u,open}}. \quad (4)$$

We note that for any PV configuration, $PAR_u(AV) \leq PAR_u(open)$ which implies that $0 \leq Y_{PAR} \leq 1$. Similarly, PV energy yield ratio is defined as:

$$Y_{PV} = \frac{\text{PV energy/unit farm area in AV configuration}}{\text{PV energy/unit farm area in standard configuration}} \quad (5)$$

A world map of the $Y_{PV}$ for fixed and sun tracking bifacial arrays based on our modeling approach have previously been reported [21-23]. Similar world maps for $Y_{PAR}$ can easily be obtained.

Finally, $LPF$ is calculated as:

$$LPF = Y_{PV} + Y_{PAR} \quad (6)$$

The individual components of $LPF$ can be optimized to maximize $LPF$ subject to the constraint of maximum allowed loss in PAR and the crop productivity. In general, the AV modules are installed at a reduced spatial density with $p/h$ ratio being 2 or 3 times larger than a standard solar PV farm and $LPF$ can range between $1 - 2$. Incidentally, it is instructive to compare $LPF$ with $LER$ which is commonly used to evaluate land productivity as [24]:

$$LER = Y_{PV} + Y_{crop(AV)}/Y_{crop(open)} \quad (7)$$

where $Y_{crop(AV)}$ and $Y_{crop(open)}$ are the *measured* crop yields obtained for AV farm and conventional farm under open sun, respectively. For a typical crop behavior, $Y_{crop} \propto PAR_u$, for which $LPF$ and $LER$ are expected to be highly correlated. This provides a great opportunity of using $LPF$ for crop-specific optimization of the PV array density, orientation, and the mobile tilt algorithms without the need for measuring the actual crop yields. $LPF$ based optimization could then be combined with other important factors, such as the relative economic importance of food vs. energy, levelized cost of electricity (LCOE) and crop revenue, *etc.* to assess, design, and predict the techno-economic performance of an AV farm.

III. RESULTS

The results are obtained for bifacial AV systems with c-Si PV modules of height ($h$) that are mounted at an elevation ($E$) above the ground and are separated by row-to-row pitch ($p$) in either N/S or E/W faced orientations. The albedo reflection ($R_A$) is taken as 0.25 for simplicity although the practical albedo varies with the type of vegetation and time of the day. As an illustrative example, all the simulations are performed for Lahore, Pakistan unless otherwise specified.

*A. Fixed tilt agrivoltaic bifacial farms*

Most of the existing fixed tilted AV farms today rely on N/S faced monofacial panels. With increasing commercial viability of bifacial panels, newer options include vertical farms oriented along the E/W faced direction as well as standard N/S faced fixed tilt PV systems. Here we explore the relative performance of N/S tilted vs. E/W vertical bifacial PV configurations. The spatially averaged PAR on the ground as a function of time of the day is shown in Fig. 5 for open farm and under different $p/h$ densities for N/S faced (left column) and E/W faced vertical (right column) PV arrays across various months. For both orientations, PAR reduces with increasing array density, as expected. An important difference in the trends for the two orientations is that for E/W, PAR reduction is more prominent during early morning and closer to sunset, whereas for N/S arrays, PAR reduces more around the middle part of the day. As a result, E/W vertical arrays tend to meet the $PAR_{th}$ around noon but remains lower than the threshold during early morning and late afternoon. The opposite is true for the N/S faced arrays. The implication of this difference in the temporal behavior for the crop yield remains an open question and should be explored in future studies.

Fig. 6 shows the monthly $Y_{PAR}$ for the N/S and E/W faced arrays. $Y_{PAR}$ decreases as a function of increasing array density, as expected. The relative decrease in $Y_{PAR}$ as a function of array density is however highly dependent on the crop's $PAR_{th}$. While $Y_{PAR}$ for lettuce remains above 80% for full or lower array densities, turnip and corn show significantly lower $Y_{PAR}$ at full or higher array densities. An important difference between the trends for N/S and E/W faced arrays is a higher seasonal variation of $Y_{PAR}$ for the former which is more prominent for the shade sensitive crops. A peak centered around the month of May is observed for turnip and corn which gradually lowers as we move towards winter months. The seasonal variation is less prominent for lower array density ($p/h > 2$). The seasonal variation of crop-specific $Y_{PAR}$ should be considered to assess the crop yield while comparing various PV array topologies.

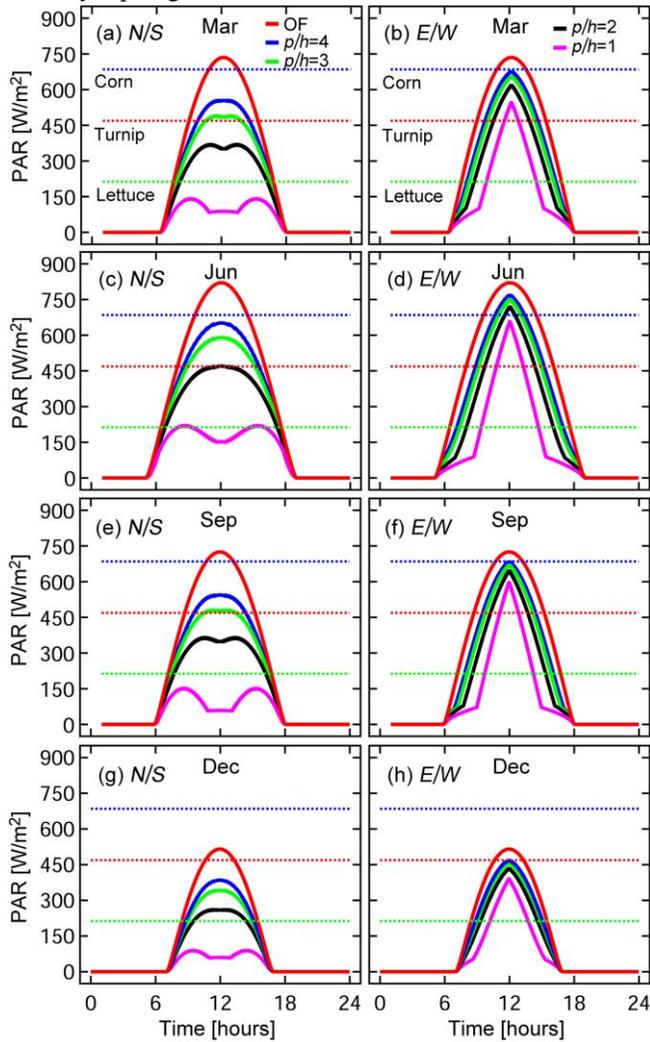

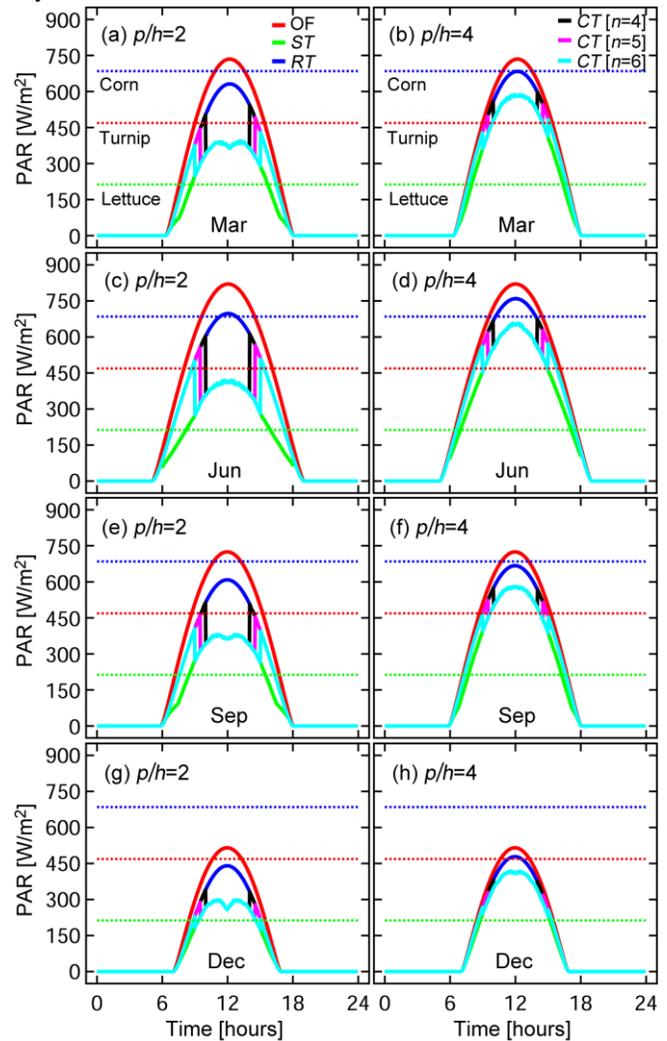

Fig. 6 Monthly $Y_{PAR}$ for three different crops under N/S (top row) and vertical E/W (bottom row) faced fixed tilt PV farms at different PV array densities.

Fig. 5 PAR available to crops under open and varying fixed tilt PV array density from half to double for (a) N/S faced arrays (left column), and E/W faced vertical arrays (right column) for various months. Horizontal lines represent the $PAR_{th}$ for lettuce, turnip, and corn.

Fig. 7 PAR available to crops under open and single axis tracking PV arrays at (a) full density (left column), and (b) half density (right column) for various months. Horizontal lines represent the $PAR_{th}$ for lettuce, turnip, and corn.

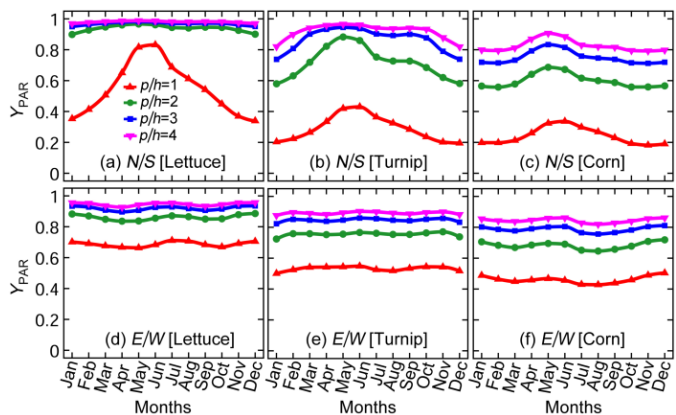

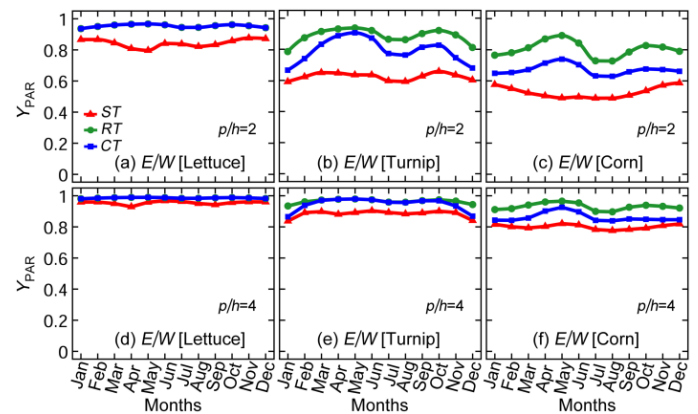

Fig. 8 Monthly $Y_{PAR}$ for three different crops under standard (ST),

reverse (RT), and customized (CT) single axis tracking schemes at full (top row) and half (bottom row) density PV arrays.

### B. Tracking agrivoltaic bifacial farms

Fig. 7 show the spatially averaged daily PAR variations under open sun and for different single axis tracking schemes at $p/h = 2$ and $p/h = 4$, respectively, for the months of March, June, September, and December. The relative PAR difference between ST and RT is smaller towards early morning and late afternoon but grows larger around noon. For CT, the tracking scheme switches from RT to ST for a few hours around noon. The customization of this switching for a given crop allows a great flexibility to maximize the $PAR_{th}$ requirement for a given month. Although PAR reaching to the ground decreases when switching from RT to ST, values close to $PAR_{th}$ could still be achievable due to high sunlight intensity closer to noon. For full array density, a noticeable difference in PAR between RT and ST is also observed in early morning and late afternoon. For half density, the overall difference in PAR between the various tracking is relatively small as compared to that for the full density. Moreover, during the early morning and late afternoon, difference in PAR between the tracking schemes is negligible for the half density.

Fig. 8 shows $Y_{PAR}$ for ST, RT, and CT ($n = 2$) schemes at $p/h = 2$ and $p/h = 4$ for the three crops across all months. It can be noted that ST results in limiting $Y_{PAR}$ below 60% for the full density array except for lettuce. Using RT, $Y_{PAR}$ recovers to $\geq 80\%$ for all crops except for a few summer months where corn shows a slightly lower $Y_{PAR}$. For half density arrays, ST provides $Y_{PAR} \geq 80\%$ for all three crops across all months. The customization of CT can optimize $Y_{PAR}$ within the margins available between RT and ST which widens for the full density as compared to the half density.

Fig. 9 compares the annual $Y_{PV}$, $Y_{PAR}$, and $LPF$ for all fixed and mobile PV schemes under study at full and half densities of the PV arrays. The lowest and highest $Y_{PV}$ is for the RT and ST tracking schemes while the opposite is true for $Y_{PAR}$. The trend for $LPF$ is closer to that of $Y_{PV}$. It should be noted that the relative difference between $Y_{PV}$, $Y_{PAR}$ and $LPF$ for E/W vs. N/S faced fixed tilt panels is not significant at half density. The maximum $LPF$ is about 1.6 and 1.9 for half and full density arrays, respectively.

### C. Model comparison with experimental data:

Fig. 10 shows a comparison of $Y_{PAR}$ computed for winter wheat and potato with the relative crop yield data from a field experiment [25] on an agrivoltaic farm at Heggelbach, Germany having 2.5 hectares area and 194.4 $KW_p$ installed capacity. $PAR_{th}$ of 1000 μmol /$m^2 - s$ and 504 μmol /$m^2 - s$ for winter wheat and potato, respectively [26, 27] are used in the model to compute $Y_{PAR}$ and typical meteorological data for the field location is used [19]. The relative crop yield in the experiment is determined by taking the ratio of crop yields for AV to that under open sun. The panel density, azimuth, and tilt angles are $p/h = 2.8$, $52°$ from south, and $20°$ respectively for both model and the experiment. The range of the modeled $Y_{PAR}$ in Fig. 10 represents monthly variation which have an increase from winter to summer months (as seen in Fig. 5). For potato, the experimental crop yield lies within the range of the modeled $Y_{PAR}$. For winter wheat, the relative crop yield slightly exceeds

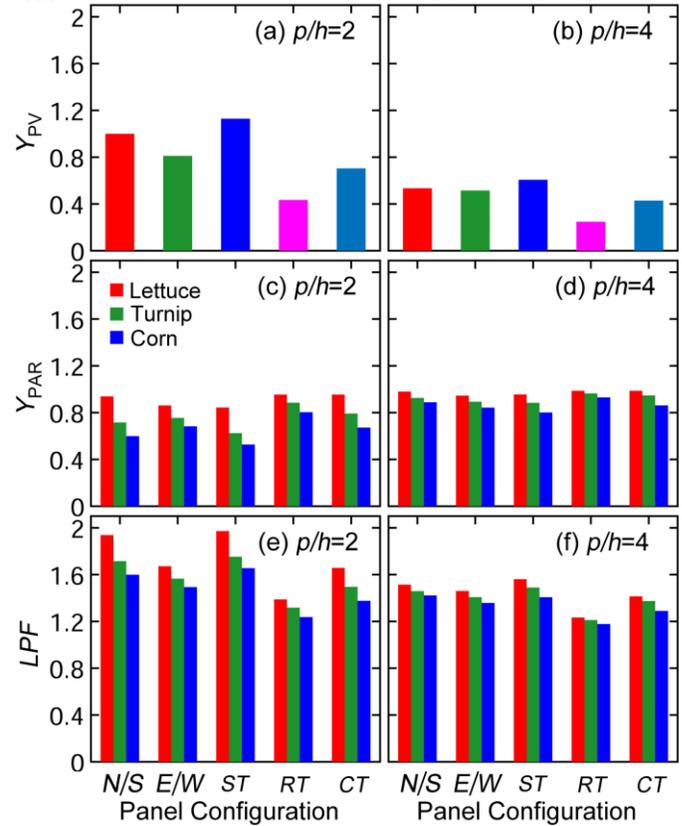

Fig. 9 (a) PV energy yield ($Y_{PV}$), (b) Useful PAR yield ($Y_{PAR}$), and (c) Light productivity factor ($LPF$) for various fixed tilt and mobile panel configurations. The left and right columns show results for full density ($p/h = 2$) and half density ($p/h = 4$) respectively.

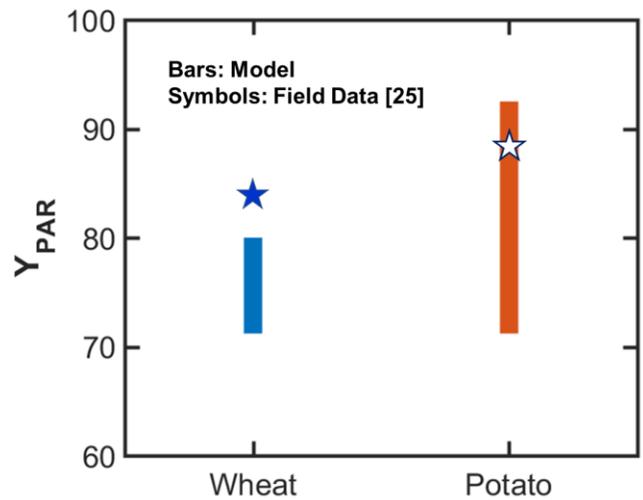

Fig. 10 Comparison of $Y_{PAR}$ with relative crop yield from an agrivoltaic field study [25].

the modeled $Y_{PAR}$ which may indicate a relatively enhanced radiation use efficiency for the field crop under AV shades.

*D. AV farm design under $Y_{PAR}$ constraint*

While $LPF$ can be maximized by increasing the array density, the resulting $Y_{PAR}$ may not always be acceptable from tolerance (turnip), N/S faced array can provide $LPF$ of 1.5 – 1.9 with ¾ array density for winter and full array density for spring, respectively. The E/W vertical array provides $LPF$ of 1.4 – 1.5 the crop yield perspective. One way to explore the design options and the relative system performance is to specify a constraint on the minimum acceptable $Y_{PAR}$. Fig. 11 shows $Y_{PAR}, Y_{YPV}$, and $LPF$ for N/S (left column) and E/W (right column) fixed tilt schemes for the array densities (full, half or ¾) which can maximize $LPF$ while satisfying the constraint of minimum $Y_{PAR}$ of ~80% for the crop across all seasons. For the shade tolerant crop (lettuce), $LPF \approx 1.9$ can be achieved for both PV orientations at full array density. For moderate shade for the corresponding seasons at ¾ density. For the case of highly shade sensitive crop (corn), both N/S and E/W arrays provide $LPF$ of ~1.3 across all seasons at half and ¾ array density, respectively.

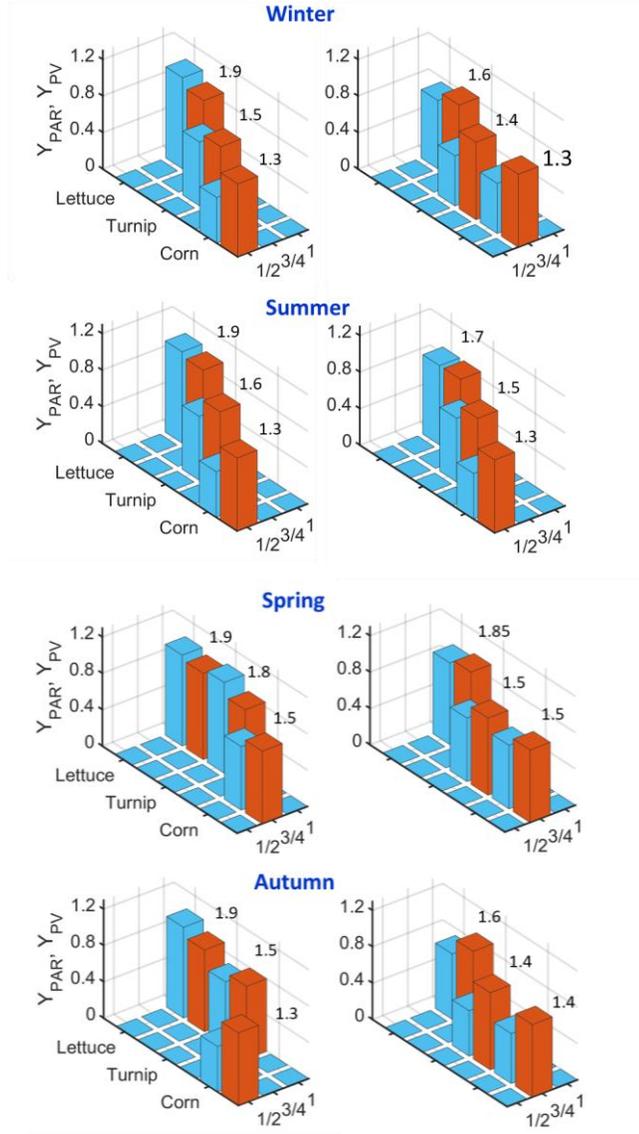

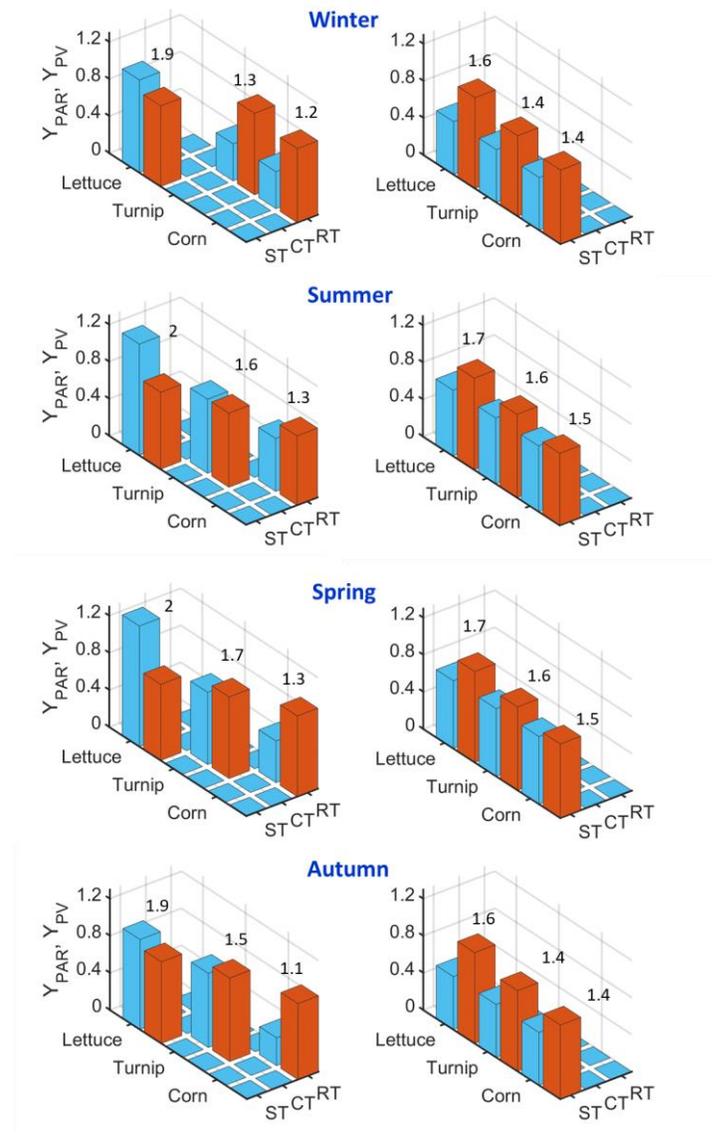

Fig. 11 $Y_{PAR}, Y_{PV}$, and $LPF$ for the N/S (left column) and E/W (right column) fixed tilt PV schemes for lettuce, turnip, and corn across various seasons. Only those crop/PV scheme combinations are shown which satisfy the constraint of minimum $Y_{PAR}$ of ~80%. The red and blue colors are for $Y_{PV}$ and $Y_{PAR}$ respectively. The values written on top of each set of $Y_{PV}$ and $Y_{PAR}$ bars are the $LPF$ of the corresponding crop/PV scheme combination.

Fig. 12 $Y_{PAR}, Y_{PV}$, and $LPF$ for single axis tracking schemes (ST, CT, and RT) at half (left column) and full (right column) array density for lettuce, turnip, and corn across various seasons. Only those crop/PV scheme combinations are shown which satisfy the constraint of minimum $Y_{PAR}$ of ~80%. The red and blue colors are for $Y_{PV}$ and $Y_{PAR}$ respectively. The values written on top of each set of $Y_{PV}$ and $Y_{PAR}$ bars are the $LPF$ of the corresponding crop/PV scheme combination.

Fig. 12 shows $Y_{PAR}, Y_{YPV}$, and $LPF$ for single axis tracking schemes (ST, CT, and RT) at half (left column) and full (right column) array density where the $LPF$ is maximized under the constraint of minimum $Y_{PAR}$ of ~80% for the crop across various seasons. For the shade tolerant crop (lettuce), ST can be implemented for both half and full array densities resulting in $LPF$ of $1.8 - 2$ and $1.5—1.9$, respectively. For the case of highly shade sensitive crop (corn), ST can still be implemented for half density arrays, but the full density array requires RT. It can be noted that for the most shade sensitive crop, ST at half array density results in better $LPF$ as compared to RT at full density. Finally, the CT provides an opportunity to maintain the required $Y_{PAR}$ for turnip at full array density when ST fails due to its lower $Y_{PAR}$.

## IV. Conclusions

In this paper, we have proposed a simple metric to characterize the efficacy of sunlight sharing between PV modules and the crops for agrivoltaic systems. The approach is based on evaluating the agrivoltaic energy yield and the daily PAR utilized in crops' photosynthesis, relative to the standalone PV and open crops, respectively. Light productivity factor ($LPF$) is introduced as a metric to indicate the overall efficacy of sunlight sharing between the crops and the PV modules. The $LPF$ varies with the crop's sensitivity to the shade and the system parameters such as the spatial density of PV arrays, panel tilt/orientation, and the tracking scheme, *etc*. We have explored these design parameters taking lettuce, turnip, and corn as representative crops for low, moderate, and high shade sensitivity, respectively. It is shown that although high density arrays and standard sun tracking can maximize $LPF$, the PAR requirement for the crop can be drastically reduced. To ensure an acceptable crop yield while optimizing $LPF$, a minimum constraint on the useful daily PAR reduction for the crop should be enforced. We provide an illustrative example of exploring the system design under a constraint of at least ~80% useful daily PAR for the crops relative to an open system and present the following key points:

- Crops that are highly shade tolerant allow $LPF$ of ~2, while the crops having moderate and high shade sensitivity permit $LPF$ in the range of 1.2 to 1.8, respectively, across various seasons.
- Full array density may be used for shade tolerant crop across all seasons, whereas half or ¾ array density is required for moderate to highly shade sensitive crops, respectively for various seasons. These results are consistent with empirical results reported in the field experiments [4, 25].
- For the fixed tilt PV arrays that include N/S faced optimally tilted and E/W faced vertical bifacial configurations, the relative difference in the performance is negligible for the half density arrays but becomes significant at full array density for which N/S faced arrays show better LPF.
- E/W faced vertical PV stands out with a least variation in the seasonal crop performance and may be the preferrable fixed tilt scheme since it can provide other important benefits such as low elevation mounting, easy movement of farm equipment, and reduced soiling loss.
- For full PV array density, the standard PV tracking can provide $LPF$ ~2 for the shade tolerant crops but customized or reverse tracking schemes are required for moderate and shade sensitive crops providing $LPF$ 1.2—1.6 depending on the season. Reducing the PV array density to half allows implementing the standard solar tracking with the $LPF$ of ~1.4 – 1.7 for both shade sensitive and tolerant crops respectively across all seasons.
- For the most shade sensitive crop, ST at half array density provides better $LPF$ (~1.5) as compared to RT at full density which permits $LPF$~1.2.

The approach and performance metrics presented in this paper to characterize agrivoltaic systems can be valuable for an initial technology assessment and design optimization, comparative analysis of agrivoltaic system configurations, and an early indication of the crop-energy yields. In particular, this work highlights that $LPF$ formulism offers an improved design of experiments. When combined with field study, our approach provides an opportunity to develop deeper insights into the farm productivity by evaluating relative difference between $LPF$ and $LER$. Although the $LPF$ approach is applied to single crops, its multi-cropping generalization is easy to develop. Finally, we would like to emphasize that the region-specific complexity of policy for food, energy, *etc*. is important in determining the optimal farm design. For example, the relative socio-economic significance of food vs. energy and relative LCOE vs. standard PV farms are important considerations. In principle, the agrivoltaic price-performance ratio could be better because of the agricultural production serving as the second resource of revenue in addition to the electricity. The future work on this topic should quantify these region-specific practical considerations for the techno-economic design and assessment of agrivoltaic technologies.

## V. References

1. Rehbein, J.A., et al., *Renewable energy development threatens many globally important biodiversity areas.* Global change biology, 2020. **26**(5): p. 3040-3051.
2. Moore-O'Leary, K.A., et al., *Sustainability of utility-scale solar energy–critical ecological concepts.* Frontiers in Ecology and the Environment, 2017. **15**(7): p. 385-394.
3. Hernandez, R.R., et al., *Techno–ecological synergies of solar energy for global sustainability.* Nature Sustainability, 2019. **2**(7): p. 560-568.
4. Barron-Gafford, G.A., et al., *Agrivoltaics provide mutual benefits across the food–energy–water nexus in drylands.* Nature Sustainability, 2019. **2**(9): p. 848-855.
5. Dinesh, H. and J.M. Pearce, *The potential of agrivoltaic systems.* Renewable and Sustainable Energy Reviews, 2016. **54**: p. 299-308.
6. Dupraz, C., et al., *Combining solar photovoltaic panels and food crops for optimising land use: Towards new agrivoltaic schemes.* Renewable energy, 2011. **36**(10): p. 2725-2732.

7. Goetzberger, A. and A. Zastrow, *On the coexistence of solar-energy conversion and plant cultivation.* International Journal of Solar Energy, 1982. **1**(1): p. 55-69.
8. Malu, P.R., U.S. Sharma, and J.M. Pearce, *Agrivoltaic potential on grape farms in India.* Sustainable Energy Technologies and Assessments, 2017. **23**: p. 104-110.
9. Amaducci, S., X. Yin, and M. Colauzzi, *Agrivoltaic systems to optimise land use for electric energy production.* Applied energy, 2018. **220**: p. 545-561.
10. M. H. Riaz, H.I., R. Younas, M. A. Alam and N. Z. Butt, *Module Technology for Agrivoltaics: Vertical Bifacial Versus Tilted Monofacial Farms* IEEE Journal of Photovoltaics 2021.
11. Marrou, H., et al., *Microclimate under agrivoltaic systems: Is crop growth rate affected in the partial shade of solar panels?* Agricultural and Forest Meteorology, 2013. **177**: p. 117-132.
12. Sekiyama, T. and A. Nagashima, *Solar sharing for both food and clean energy production: performance of agrivoltaic systems for corn, a typical shade-intolerant crop.* Environments, 2019. **6**(6): p. 65.
13. Valle, B., et al., *Increasing the total productivity of a land by combining mobile photovoltaic panels and food crops.* Applied energy, 2017. **206**: p. 1495-1507.
14. Hassanpour Adeh, E., J.S. Selker, and C.W. Higgins, *Remarkable agrivoltaic influence on soil moisture, micrometeorology and water-use efficiency.* PloS one, 2018. **13**(11): p. e0203256.
15. Othman, N.F., et al., *Modeling of Stochastic Temperature and Heat Stress Directly Underneath Agrivoltaic Conditions with Orthosiphon Stamineus Crop Cultivation.* Agronomy, 2020. **10**(10): p. 1472.
16. Carriedo, L.G., J.N. Maloof, and S.M. Brady, *Molecular control of crop shade avoidance.* Current opinion in plant biology, 2016. **30**: p. 151-158.
17. Sun, X., et al., *Optimization and performance of bifacial solar modules: A global perspective.* Applied energy, 2018. **212**: p. 1601-1610.
18. Stein, J.S., et al. *PVLIB: Open source photovoltaic performance modeling functions for Matlab and Python.* in *2016 ieee 43rd photovoltaic specialists conference (pvsc).* 2016. IEEE.
19. Zhang, T., et al. *A global perspective on renewable energy resources: NASA's prediction of worldwide energy resources (power) project.* in *Proceedings of ISES World Congress 2007 (Vol. I–Vol. V).* 2008. Springer.
20. Orgill, J. and K. Hollands, *Correlation equation for hourly diffuse radiation on a horizontal surface.* Solar energy, 1977. **19**(4): p. 357-359.
21. Khan, M.R., et al., *Vertical bifacial solar farms: Physics, design, and global optimization.* Applied energy, 2017. **206**: p. 240-248.
22. Patel, M.T., et al., *Optimum design of tracking bifacial solar farms--A comprehensive global analysis of next-generation PV.* arXiv preprint arXiv:2011.01276, 2020.
23. Patel, M.T., et al., *A worldwide cost-based design and optimization of tilted bifacial solar farms.* Applied Energy, 2019. **247**: p. 467-479.
24. Mead, R. and R. Willey, *The concept of a 'land equivalent ratio'and advantages in yields from intercropping.* Experimental Agriculture, 1980. **16**(3): p. 217-228.
25. Schindele, S., et al., *Implementation of agrophotovoltaics: Techno-economic analysis of the price-performance ratio and its policy implications.* Applied Energy, 2020. **265**: p. 114737.
26. Austin, R., *Prospects for genetically increasing the photosynthetic capacity of crops.* 1990.
27. Tazawa, S., *Effects of various radiant sources on plant growth (Part 1).* Japan Agricultural Research Quarterly, 1999. **33**: p. 163-176.